\documentclass[fleqn,twoside]{article}
\usepackage{espcrc2}

\usepackage{graphicx}
\usepackage[figuresright]{rotating}

\newcommand{\AmS}{{\protect\the\textfont2
  A\kern-.1667em\lower.5ex\hbox{M}\kern-.125emS}}

\title{Quantum-Gravity phenomenology and high energy particle propagation}

\author{R. Aloisio\address[LNGS]{INFN - Laboratori Nazionali del 
Gran Sasso, SS. 17bis\\
Assergi (L'Aquila) - Italy},
P. Blasi\address[FI]{INAF - Osservatorio Astrofisico di Arcetri, 
Largo E. Fermi 5\\
50125 Firenze - Italy},
A. Galante\addressmark[LNGS]\address[dipaq]{Dipartimento di Fisica, 
Universit\`a di L'Aquila, Via Vetoio\\
67100 Coppito (L'Aquila) - Italy}\thanks{Talk presented by A. Galante},
P.L. Ghia\address[cnr] { CNR - IFSI, Sezione di Torino, Corso Fiume 4, 
10133 Torino - Italy}\address[tor]{INFN - Sezione di Torino, 
Via P. Giuria 1, 10125 Torino - Italy},
A.F. Grillo\addressmark[LNGS]}

\begin{document}

\begin{abstract}
Quantum-gravity effects may introduce relevant consequences for the 
propagation and interaction of high energy cosmic rays particles.
Assuming the space-time foamy structure results 
in an intrinsic uncertainty of energy and momentum of particles,
we show how low energy (under GZK) observations can provide
strong constraints on the role of the fluctuating space-time structure.
\vspace{1pc}
\end{abstract}

\maketitle

\section{Introduction}

Already few years after the first theoretical study of the possible 
absorption of UHE protons on the low energy thermal photon background 
\cite{GZK}
the experimental evidence of ultra GZK cosmic rays was under debate.
This situation was the starting point for some speculative work
on possible new physics to explain an unexpected CR spectrum 
above $\sim 10^{19} eV$. To our knowledge the first authors
to recognize such possibility where Kirzhnits and Chechin in
1971 \cite{K&C}. In their pioneering work in fact they wrote 
{\it ``Primary protons with energy above $5\cdot 10^{19} eV$ are expected to
be strongly slowed down by the interaction with the background
thermal radiation. However, no break is observed in the CR
spectrum in this region. It is of course premature in this
circumstances....''} and the key observation was
{\it ``The point is that the primary
protons have a uniquely large Lorentz factor $\gamma > 5 10^{10}$
larger by many order of magnitudes than in any other experiment..''}.
With these premises 
they proposed a modified theory to introduce small violations
in the dispersion relation of particles at sufficiently high 
energies in such a way to
account for the absence of the so-called GZK feature in the spectrum.

More than 30 years after the experimental situation is still 
unclear: present operating experiments are AGASA \cite{AGASA}
and HiRes \cite{Hires}, and they do not provide strong evidence either 
in favor or against the detection of the GZK feature \cite{demarco}. 
A substantial increase in the statistics of events, as expected 
with the Auger project \cite{Auger} is expected to
clarify the scenario in one or two more years.

\section{From Quantum Gravity to Lorentz invariance breaking}

The theoretical approach to possible violations to standard
($i.e.$ Lorentz and Poincar\'e invariant) physics
changed substantially in last years.
It is not any more a simple exotic possibility to interpret unclear
experimental data but
there is a growing feeling that such violations will be a necessary
ingredient to properly describe phenomena at very small 
distance scales.

The most ambitious program tries to merge the typical quantum
behavior and the General Relativity in a well defined theory
of Quantum Gravity (QG).
Several attempts to construct a model for QG
have been done. They basically share a new interpretation for
space-time: it is no more a given background for physical objects
but, more properly, a derived concept itself.
It is expected to be characterized by a typical length scale
(the Planck length $l_P\simeq 10^{-33}$ cm is a natural candidate)
where its structure 
becomes dominated by quantum fluctuations and basically
undefined (at least for our present  capabilities to describe it).
This attempts include Loop QG, some string-based model and the 
space-time foam approach (the latter is indeed the older one
and traces back to '60s \cite{wheeler57}).

Apart from those first principles constructions other models have
been proposed as effective theories that should try to catch some
of the possible new QG physics at small but still super Planckian
length scales. They basically include the large family of models 
based on Quantum deformed Poincar\'e Groups as well as the Doubly 
Special Relativity construction(s).

All these approaches are very different and they predict some
modification of basic physical principles. The following is
a non-exhaustive cumulative list of the different possibilities.
The first is the possibility of modification of Poincar\'e and
Lorentz symmetries. Depending on the specific model they can 
still be exactly realized as well as 
explicitly broken (introducing a preferred reference frame)
or kept but in a deformed way. Also the energy-momentum
relation can be modified including extra terms that can be of
fixed or stochastic nature.  
Generally a new invariant physical scale ($l_p$ or the Planck
energy $E_p$) is introduced and this scale can (eventually)
coexist with the standard invariant: 
the low energy light velocity $c$.
Other possible effects are indetermination in position and/or
momentum measurements due to the fluctuating nature of the 
space-time structure and the appearance of new non-linear 
composition laws for energy and momentum of multiparticle 
states $i.e.$ $P_{tot} \neq \sum_i P_i$.

Many of these possibilities have been investigated
trying to find possible experimental signatures for new
physics even at energy scales much smaller the $10^{28}$ eV
that correspond to the Plank energy.

Astroparticle physics is a privileged arena for such studies 
both for the availability of very energetic particles
and for the possibility to consider their motion along large 
(cosmological) distances.
Among the others the large distance propagation of photons
with energy dependent velocity \cite{am97,biller99} 
and modifications induced in 
the standard synchrotron radiation emission process have been
considered to put limits on possible Lorentz Invariance (LI) 
breaking \cite{jac03,ellis03,castorina}.

Another interesting possibility to test such models is to
consider  physical processes with a kinematic energy threshold, 
which is in turn very sensitive to the smallest violations of LI.
This is the case for UHECRs and TeV gamma rays.
UHECRs are expected to suffer severe 
energy losses due to photopion production off the photons of the cosmic 
microwave background (CMB), and this should suppress the flux of particles 
at the Earth at energies above $\sim 10^{20}$ eV, the so called GZK feature. 
Super-TeV energy photons from sources at cosmological distances
are expected to undergo electron-positron production in interactions
with low energy photons of the far infra red background (FIRB) and 
CMB.

In both cases a very large $\gamma$ factor is involved in moving
from the laboratory to the center of mass reference frame.
The sharply defined thresholds can be substantially shifted 
(or even disappear) if a small LI breaking term is introduced
giving potential for investigation in this field.
The new phenomena, if present,  
should show up in modification of expected UHECRs
spectrum.

Some authors \cite{cam,colgla,berto}
have invoked possible violations of LI as a plausible explanation to
some puzzling observations related to the detection of ultra high energy
cosmic rays (UHECRs) with energy above the GZK feature, 
and to the unexpected shape of the spectrum of photons with super-TeV 
energy from sources at cosmological distances.
 
Both types of observations have in fact 
many uncertainties, either coming from limited statistics of very rare events,
or from accuracy issues in the energy determination of the detected 
particles, and most likely the solution to the alleged puzzles will come from 
more accurate observations rather than by a violation of fundamental 
symmetries.
 
For this reason, from the very beginning we proposed \cite{noi1} that 
cosmic ray observations should be used as an ideal tool to constrain 
the minuscule violations of LI, rather than as evidence for the need 
to violate LI.

We adopt  some reasonable choice to parametrize the LI violations
predicted by QG models, consider the theoretical consequences 
and compare with experimental data.
If the features in the spectrum related to the processes thresholds 
are indeed found this will provide limits on LI violation scale.
If such features are absent this will allow us to reject some
models but, for the moment, not to prove the existence of 
LI breaking new phenomena.

\section{Fixed violations}

The recipes for the violations of LI generally consist of requiring an 
{\it explicit} modification of the dispersion relation of high energy 
particles.
This modification is an effective way to describe 
their propagation in the ``vacuum'', now affected by 
quantum gravity (QG) phenomena. 
This effect is generally parametrized by introducing 
a mass scale $M$, expected to be of the order of the Planck mass,
that sets the scale for QG to become effective. 

Without referring to any specific model, we write a modified 
dispersion relation obeying the following postulates:

1)
violations are universal, $i.e.$ do not depend on
particle type;

2)
violations preserve rotational invariance;

3) 
violations are an high energy phenomenon, vanishing at  low 
momenta.
\\
\noindent
With these conditions we write the following expression:

\begin{equation}
E^2-p^2=m^2+p^2f(p/M)
\label{disp_rel}
\end{equation}
This deformed dispersion relation has been proposed by several
authors \cite{luk98,ellis99,DSR1,DSR2} and is the most popular in the
literature.

Just for completeness we
note that another possibility compatible with the dimensional
analysis exists: it refers to the so called conformal models of
LI breaking and was considered by Kirzhnits and Chechin in their
paper. It accounts to introduce the extra (respect to the standard
case) term proportional to the particle mass squared instead that
to $p^2$. When considering thresholds modifications this last 
possibility gives no detectable effects for UHECRs propagation
if $M$ is the Planck mass.

The standard way to proceed is to expand the last term in rhs
of (\ref{disp_rel}) and this, at lowest order, gives a term of the
form 
\begin{equation}
\pm (p/M)^\alpha
\label{extra}
\end{equation}
where $\alpha$ is model dependent.
To get a quick result and some physical insight
we can argue that, for massive particles, the above extra term in
dispersion relation becomes relevant for the kinematics of
particle interactions
when its modulus is comparable with the particle squared mass.
For the protons ($i.e.$ for the GZK case) 
we get immediately the following numbers for the critical momentum
$p_c$ where we may expect changes (in the following formula we fix
$M$ to the Planck mass value):

\begin{eqnarray}
\alpha=1 &\to & p_c=(m_p^2 M^2)^\frac{1}{3}\simeq 10^{15} eV << M \nonumber\\
\alpha=2 &\to & p_c=(m_p^2 M^2)^\frac{1}{4}\simeq 10^{18} eV << M \nonumber
\end{eqnarray}

In both case we see that the value of $p_c$ is much smaller than the
Planck mass scale. This gives a first indication that if we modify
the dispersion relation with terms related to some scale (the Planck
mass in our case),
the resulting particle kinematics can indeed be sensitive to such 
changes already at much lower energy scales. In other words  
possibly we do not need Planck scale experiments to detect effects
related to new physics at Planck scale.

A detailed calculation of photopion and $e^+e^-$ threshold production 
for high energy protons and photons interacting with low energy
background photons has been carried out \cite{noi1}. In this calculation
the conservation of total energy and momentum of incoming and outcoming
particles is assumed. 

If the total energy and momentum of multiparticle states are calculated
as usual (just the sum of the contribution of each particle) and we assume
that the scale parameter $M$ is the Planck mass we find that the 
GZK feature could be absent (the threshold goes to infinity) when
we consider the minus sign in (\ref{extra}) or, for positive
sign, shifted downward by five ($\alpha=1$) or one ($\alpha=2$) order 
of magnitude respect to the standard case.

The same calculation can be done in the framework of Doubly Special 
Relativity. In this case the theory is constructed in such a way 
that the relativity principle is still valid: no privileged 
reference system exists.
The (non linear) deformed boost in
momentum space require
a change in the dispersion relation as the one previously considered
but also a different definition of total energy and momentum in multiparticle
states. For the DSR1 \cite{DSR1} and DSR2 \cite{DSR2} models we have 
\cite{judes04}:
\begin{eqnarray}
E_{tot} = E_1 + E_2 - \frac{1}{2}\frac{1}{M} (p_1 p_2 + p_2 p_1) + 
O(\frac{1}{M^2}) \nonumber\\
E_{tot} = E_1 + E_2 + \frac{1}{M} (E_1 E_2 + E_2 E_1) + 
O(\frac{1}{M^2}) \nonumber
\end{eqnarray}
In this case basically no new particle processes (respect to the
standard theory) are kinematically allowed and, for the GZK case,
the momentum threshold is basically the same as in standard case
\cite{major04}.

In drawing conclusions from this kind of studies we have to keep in
mind that there are two main problems. The first is the 
poor knowledge of UHECRs sources, a piece of information needed to
predict the actual effect of photopion production on the cosmic ray
spectrum and, at the end, necessary to correctly interpret the 
experimental data. 
The second is related to the limitation of approaches based uniquely
on kinematic analysis: the present impossibility to include 
the dynamical effects of the full theory makes quantitative 
conclusions questionable (even if it seems reasonable to 
expect modifications to dynamics to be proportional to the energy scale
divided by $M$ and hence highly suppressed for physics below
GZK scale).

At the end, after a suitably parametrization of LI 
violating models and an analysis of interaction kinematics,
we conclude that it is actually premature to explain GZK
absence as a manifestation of LI violations.
On the other side, once the experimental situation will be 
clarified, it will be possible to put strong limits on new
physics $i.e.$ on the breaking parameter $M$ considered as
an independent quantity.

\section{From fixed to stochastic violations}

One can ask if the above is the only possibility to introduce
some remnant of QG effects in low energy particle behavior.
To answer this question we first note that it is generally
believed that coordinate measurements cannot be performed 
with precision better some quantity related to the 
Planck distance (time): $\delta l \ge l_p (l/l_p)^\eta$ (where 
$\eta=0, 1/2, ...$ depending on the particular model),
since such a measurement would result in the production of a 
black hole.

We consider the $\eta=0$ case and argue that such indetermination
can be seen as the result of a fluctuating metric tensor $g_{\mu\nu}$.
The metric tensor, has felt by a propagating particle
of wavelength $\lambda$, can be written as a ``standard'' term plus
a variation $\delta g_{\mu\nu}$ 
that should account for the quantum fluctuations
of space-time. If we stick again to the postulates of previous section
we can write $\delta g_{\mu\nu} = h_{\mu\nu} (l_p/\lambda)$ where
$ h_{\mu\nu}$ is a rotationally symmetric tensor of order 1.

To go from position indetermination $\delta l = l_p$ 
to momentum and/or energy fluctuation we need to assume the
validity of some sort of QG-modified De Broglie relation
(see for example \cite{kempf95})
\begin{equation}
\delta p = \delta\left\{\frac{1}{\lambda}\left[1+f\left(\frac{l_p}{\lambda}
\right)\right]\right\}\simeq p^2 l_p + O(l_p^2)
\label{Debroglie}
\end{equation}
where $\lambda$ is again the particle wavelength. We think it is safe to
assume the above relation
to have some sense at least for energies small compared to $M$.

We can conclude that energy and momentum of particles are not 
constant during the propagation: their measurement would result in
slightly different values in different space-time positions.
If we consider each particle interaction equivalent to a
$E, p$ measurement it results that kinematics becomes stochastic:
a given process has some probability to be allowed even for four momentum
below the classical threshold value.
So, in general, we can add a fluctuating term to $E, p$ and to the 
dispersion relation (the latter eventually adding a fixed extra term
like the one considered in previous section) and write
\begin{eqnarray}
\label{fluct}
E_i &=& \bar{E}_i + \alpha_i \frac{\bar{E}_i^2}{M} \nonumber\\
p_i &=& \bar{p}_i + \beta_i \frac{\bar{p}_i^2}{M} \\
E_i^2- p_i^2 &=& m_i^2 + \gamma_i \frac{\bar{E}_i^3}{M} \nonumber
\end{eqnarray}
where $i$ labels the interacting particle, $\bar p$ and $\bar E$ 
are the mean ($i.e.$ classical) values. 
$\alpha_i, \beta_i, \gamma_i$ are
random variables expected to be of order one that we assume to 
be statistically independent since the typical interaction scale
in the processes we are interested ($E\le 10^{20}$ eV) is much larger
that the QG scale $M$.
The above expressions have been motivated in various ways  
in previous papers
\cite{ford99,ng01,lieu02,noi2,camacho03}.

Assuming total energy-momentum conservation and standard composition
laws we find that a proton with energy $E > 10^{15}$ eV has $\simeq 70 \%$
probability to interact with a $\gamma_{CMB}$ thus loosing 
some of its energy.
The result is insensitive to the detail of the fluctuation $i.e.$
it is valid basically for any choice of the probability distribution 
function of  the $\alpha,\beta,\gamma$ \cite{noi2}.

Apart from possible modification of $p-\gamma_{CMB}$ and 
$\gamma-\gamma_{FIRB}$ thresholds we also considered the effect of
fluctuations on particle stability.
Indeed, due to fluctuations, a particle propagating in vacuum 
acquires an energy dependent (fluctuating) mass, which may be responsible
for kinematically forbidden decays to become kinematically allowed.

It is the case for photon decay $\gamma\to e^+ e^-$, possible
as soon as the photon has momentum $p > p_{th} \simeq 10^{13}$ eV,
and proton decay $p \to p + \pi^0$, possible for 
$p > p_{th} \simeq 10^{15}$ eV.
But the most dramatic effect would be the Vacuum Cerenkov radiation
emission from charged particles $x \to x + \gamma$.
In this case the threshold can be written as
\begin{equation}
p_{th} \simeq \left(\frac{m^2 M \omega}{\delta}\right)^{\frac{1}{4}}
\end{equation}
where $m$ is the particle mass, 
$\omega$ is the photon energy and $\delta$ is some combination
of the fluctuating coefficients of (\ref{fluct}).
Clearly $p_{th} \to 0$ if $\omega \to 0$ and this will eventually
result in a stability crisis for all charged particles \cite{noi3}.

This scenario is clearly experimentally excluded and we can go back
trying to understand if the apparently natural fluctuation picture has to be
discarded or some possible way out is still present.
We first note that the presence of fixed modifications in the
dispersion relation does not change the result. 

A complete analysis unfortunately would imply some knowledge of the
dynamical effects to answer questions like
\begin{enumerate}
\item if the particles were kinematically allowed to decay, and there
were no fundamental symmetries able to prevent the decay, would it 
take place? 

\item is the form adopted for the quantum fluctuations correct and if so,
how general is it?

\item are the various fluctuating terms really independent?

\item if in fact the form adopted for the fluctuations is correct, how
general and unavoidable is the consequence that (experimentally)
unobserved decays should take place?
\end{enumerate}

Another interesting point is to consider the combined effect of assuming
different energy-momentum conservation laws, like the
one in DSR models, and fluctuations \cite{gallou04}

There is no answer to this questions at the moment but we can say
that the QG phenomenological models considered here can already
be excluded. 

In any case the positive conclusion is that 
it is clear that UHECRs are a powerful tool to
test the LI breaking scenario either making QG phenomena detectable
or severely constraining it.

\end{document}